\documentclass[graybox]{svmult}

\usepackage{mathptmx}       
\usepackage{helvet}         
\usepackage{courier}        
\usepackage{type1cm}        
\usepackage{makeidx}         
\usepackage{graphicx}        
\usepackage{multicol}        
\usepackage[bottom]{footmisc}

\usepackage{amssymb}
\usepackage{amsmath}
\usepackage{mathtools}
\usepackage{etoolbox}
\usepackage{enumerate}

\usepackage{tikz}

\usepackage[pdfborder={0 0 0 [3 3]}]{hyperref}

\newcommand{\ZZ}{\mathbb{Z}}			
\newcommand{\NN}{\mathbb{N}}			
\newcommand{\RR}{\mathbb{R}}			
\newcommand{\isdef}{\triangleq}			
\newcommand{\abs}[1]{
	\left\lvert#1\right\rvert%
}

\newcommand{\xd}{\mathrm{d}}			
\newcommand{\xe}{\mathrm{e}}			

\newcommand{\symb}[1]{\mathtt{#1}}		
\newcommand{\upspin}{\symb{\uparrow}}		
\newcommand{\downspin}{\symb{\downarrow}}	
\newcommand{\XOR}{
	\mathbin{\mathsf{XOR}}%
}
\newcommand{\OR}{
	\mathbin{\mathsf{OR}}%
}

\newcommand{\bcell}[2]{%
	\draw[help lines] #2+(-0.4,-0.4) rectangle +(0.4,0.4);
	\ifstrequal{#1}{1}{%
		\fill #2+(-0.4,-0.4) rectangle +(0.4,0.4);
	}{}
}

\makeindex             

\begin{document}

\title*{Statistical equilibrium in deterministic cellular automata}
\titlerunning{Equilibrium in deterministic CA}

\author{Siamak Taati}
\authorrunning{S. Taati}

\institute{Siamak Taati
	\at Mathematical Institute, Leiden University, The Netherlands,
	\email{siamak.taati@gmail.com}
}
%
%
\maketitle


\abstract*{%
	Some deterministic cellular automata have been observed
	to follow the pattern of the second law of thermodynamics:
	starting from a partially disordered state, the system evolves towards
	a state of equilibrium characterized by maximal disorder.
	This chapter is an exposition of this phenomenon and
	of a statistical scheme for its explanation.
	The formulation is in the same vein as Boltzmann's ideas,
	but the simple combinatorial setup
	offers clarification and hope for generic mathematically rigorous results.
	Probabilities represent frequencies
	and subjective interpretations are avoided.
}

\abstract{%
	Some deterministic cellular automata have been observed
	to follow the pattern of the second law of thermodynamics:
	starting from a partially disordered state, the system evolves towards
	a state of equilibrium characterized by maximal disorder.
	This chapter is an exposition of this phenomenon and
	of a statistical scheme for its explanation.
	The formulation is in the same vein as Boltzmann's ideas,
	but the simple combinatorial setup
	offers clarification and hope for generic mathematically rigorous results.
	Probabilities represent frequencies
	and subjective interpretations are avoided.
}


\section{Introduction}
\label{sec:intro}
The aim of statistical mechanics is to bridge between
microscopic and macroscopic behaviour
of systems consisting of a large number of interacting components.
The prime example is a gas of particles moving and interacting according
to the laws of mechanics, giving rise to macroscopic behaviour
described in thermodynamics.
The kinetic theory of gases, initiated by Clausius, Maxwell and Boltzmann,
takes on the task of explaining the macroscopic behaviour of a gas
on the basis of its microscopic description.

The main problem in kinetic theory
is the derivation of the second law of thermodynamics
(i.e., the tendency of an isolated thermodynamic system to evolve
towards more disordered states).
Starting from a collection of particles pictured
as hard balls interacting through elastic collisions
and using a simplifying (though erroneous) statistical assumption about the number of
collisions of each type occurring in a small time interval
(the \emph{Stosszahlansatz}),
Boltzmann was able to derive a version of the second law
by showing that a certain quantity measuring disorder (Boltzmann's entropy)
increases monotonically with time and is maximized precisely when
the system is in equilibrium.

Although the second law of thermodynamics was originally formulated
for thermodynamic systems, its applicability goes beyond a system of particles
following the particular laws of (classical or quantum) mechanics.
A mathematical understanding of the precise circumstances leading
to the applicability of the more general law of tendency towards disorder
is desirable but missing.

The purpose of this chapter is to demonstrate examples of
results and experimental observations regarding
the so-called randomization behaviour in cellular automata (going back to Miyamoto, Wolfram and Lind)
that could be thought of as instances of this generalized version of the second law of thermodynamics.
Notably, neither probabilistic hypotheses
(i.e., incorporating intrinsic randomness in the model)
nor subjective interpretations
(see~\cite{Jay57}) are needed ---
probabilities enter the picture only as intuitive means of representing statistical data.
The combinatorial setting of cellular automata is simple enough that
one could attempt to find generic mathematical conditions that guarantee the applicability of
the second law.  At the same time, the rich range of behaviour among cellular automata
makes the challenge interesting and non-trivial.

The scenario is briefly as follows.
Consider a configuration that is atypical of the maximally disordered state
(so that there is a bias in the frequency of the patterns)
but is not too rigidly regular either (e.g., it is not periodic).
Over the time, a sufficiently chaotic cellular automaton
shuffles such a configuration (albeit deterministically)
in such a way that the bias gradually becomes undetectable.
More specifically, the configuration of the system
becomes more and more typical of the maximally disordered state,
up to wider and wider ranges of observation.

\section{Randomization Phenomenon: Examples}
\label{sec:examples}

\subsection{XOR cellular automaton}
\label{sec:examples:xor}
\begin{figure}
	\begin{center}
	\includegraphics[scale=1]{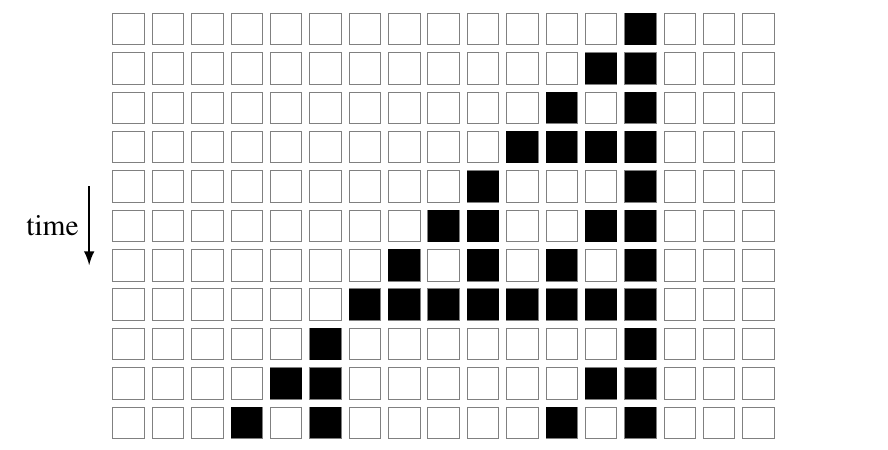}
	\caption{
		A sample trajectory of the XOR cellular automaton starting
		with a configuration consisting of
		a single $\symb{1}$ at the origin and $\symb{0}$ everywhere else.
	}
	\label{fig:xor:pascal}	
	\end{center}
\end{figure}

On the space of all bi-infinite sequences of symbols $\symb{0}$ and $\symb{1}$,
consider a transformation $T$ that maps a sequence $x$ into another sequence $Tx$
defined by $(Tx)_i \isdef x_i + x_{i+1} \pmod{2}$.
In other words, $T$ replaces each symbol with the sum (modulo~$2$)
of that symbol and its right neighbour.
The iteration of $T$ defines a dynamical system on $\{\symb{0},\symb{1}\}^\ZZ$,
which we refer to as the \emph{XOR cellular automaton}.\footnote{%
	XOR stands for ``eXclusive OR''.
}
Each sequence in $\{\symb{0},\symb{1}\}^\ZZ$ will be called a \emph{configuration} of the system.
A sample trajectory of this system is depicted in Figure~\ref{fig:xor:pascal}.

The map $T$ is continuous with respect to the product topology.
The \emph{product topology} is the topology in which two configurations are considered
``close'' if they agree on a large region around the origin.
Convergence in the product topology corresponds to site-wise eventual agreement.
Another basic property of $T$ is its translation symmetries.
Namely, if $\sigma^k$ denotes the \emph{translation} by $k$ (that is, $(\sigma^k x)_i\isdef x_{k+i}$),
then $T\sigma^k=\sigma^k T$ for every $k\in\ZZ$.
The map $T$ is also \emph{additive}, meaning
$T(x+y)=Tx + Ty$, where the addition is performed site-wise and modulo~$2$.
Although $T$ is not invertible,
it is onto and ``almost one-to-one''
in that every configuration $y\in\{\symb{0},\symb{1}\}^\ZZ$
has precisely $2$ pre-images.  Namely, choosing a symbol $x_0$ arbitrarily,
we can find, recursively, unique values for the symbols $x_i$, for $i>0$ and $i<0$,
such that $Tx=y$.

Slightly less obvious is the following \emph{balance} property of $T$:
if we choose the symbols in $x$ by independent unbiased coin flips,
the symbols in $Tx$ will also be indistinguishable from independent unbiased coin flips.
In other words, the \emph{uniform Bernoulli measure} is invariant under $T$.
To see this, take any block of $n$ consecutive symbols $b_1b_2\cdots b_n$
and consider the probability that $Tx$ takes values $b_1b_2\cdots b_n$
at positions $k$ to $k+n-1$.
There are precisely two choices of values $a_1a_2\cdots a_{n+1}$ and $a'_1a'_2\cdots a'_{n+1}$
that, if on $x$ at positions $k$ to $k+n$,
lead to the desired values $b_1b_2\cdots b_n$ on $Tx$ at sites $k$ to $k+n-1$.
Each of these two choices has probability $2^{-(n+1)}$
of appearing in independent flips of an unbiased coin.
Therefore, $b_1b_2\cdots b_n$ appears at positions $k$ to $k+n-1$ of $Tx$
with probability $2^{-n}$.

Besides the balance property,
the XOR cellular automaton has a wealth of other statistical regularities.
For instance, if $x$ is chosen according to a uniform Bernoulli measure
(i.e., with independent unbiased coin flips),
then for any $n$,
the sequence of blocks $(T^t x)_{[k,k+n)}$, $t=k, k+1, \ldots$
is independent of the block $x_{[k,k+n)}$.
It follows from the law of large numbers that,
almost surely, every pattern $a_1a_2\cdots a_n$
appears with the same frequency $2^{-n}$ in the space-time column
$(T^t x)_{[k,k+n)}$, $t=0,1,2,\ldots$.
The same sort of eventual independence holds
along any other ``space-time direction'':
for every $a\in\ZZ^+$ and $b\in\ZZ$ and a sufficiently large $t_0$,
the tilted column of blocks $(T^{at} x)_{[bt+k,bt+k+n)}$ with $t\geq t_0$
is independent of $x_{[k,k+n)}$.

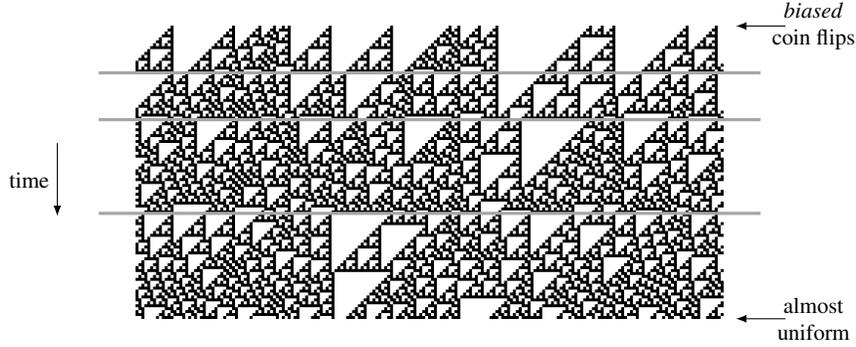
\begin{figure}
	\begin{center}
	
		\medskip
		
		\begin{tikzpicture}[scale=1.1,>=latex]
			\node[anchor=north,inner sep=0] (image) at (0,0) {%
				\includegraphics[scale=0.1375]{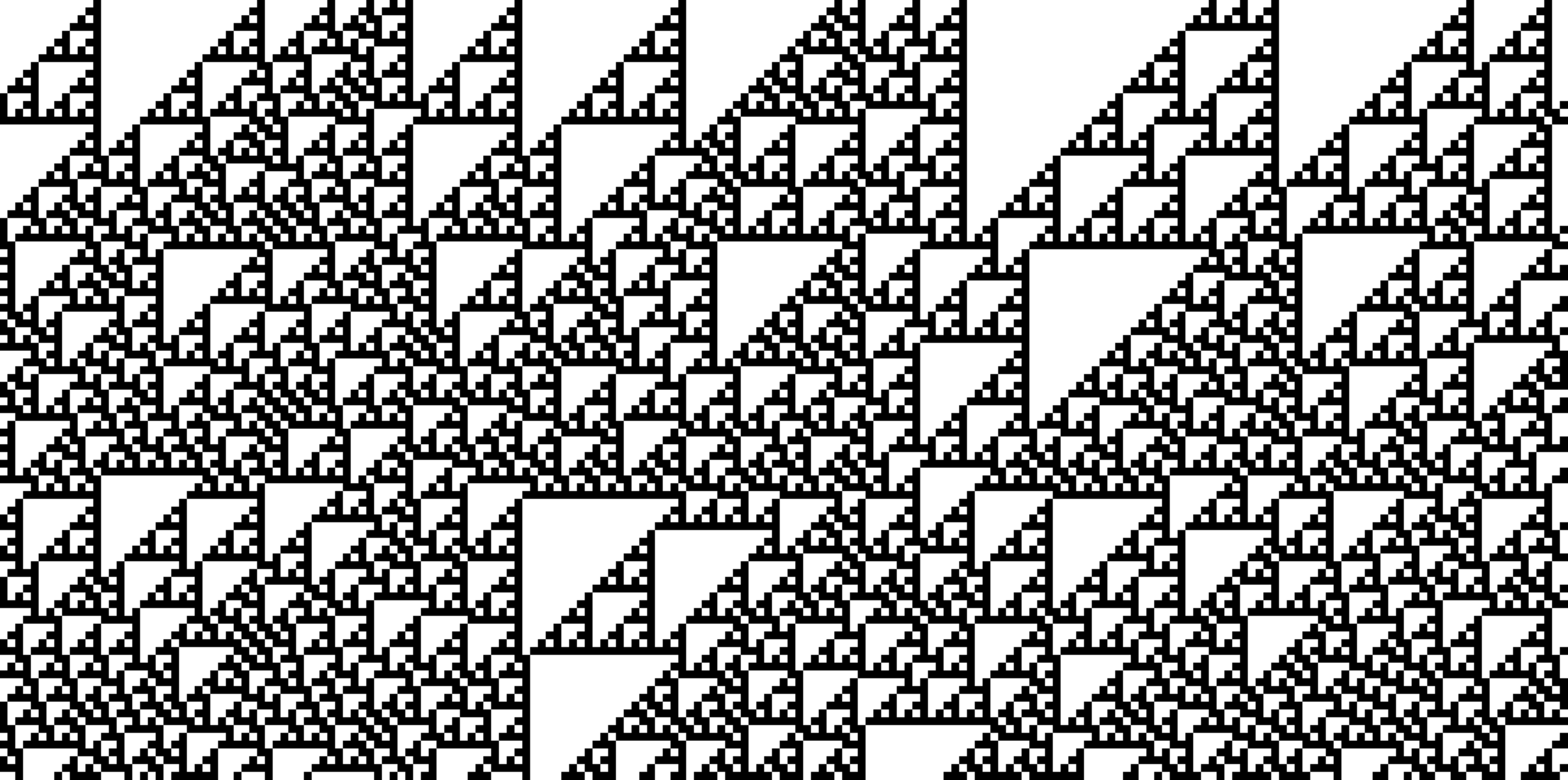}	
			};
			
			\begin{scope}[y={(image.south)}]
			
				\draw[very thick, gray!70] (-4,0.16) -- (4,0.16);
				\draw[very thick, gray!70] (-4,0.32) -- (4,0.32);
				\draw[very thick, gray!70] (-4,0.64) -- (4,0.64);
			
				\begin{scope}[overlay]
					\draw[->] (-4.5,0.4) to node[left]{\small time} (-4.5,0.65);
					
					\draw[thin,->] (4.3,0) to (3.7,0);
					\node[anchor=west] at (4,0) {%
						\parbox[t]{4em}{\centering\footnotesize\emph{biased}\\ coin flips}
					};
				
					\draw[thin,->] (4.3,1) to (3.7,1);
					\node[anchor=west] at (4,1) {
						\parbox[t]{4em}{\centering\footnotesize almost\\ uniform}
					};
				\end{scope}
			\end{scope}
		\end{tikzpicture}
		
		\medskip
	
	\caption{
		Randomization effect of the XOR cellular automaton.
		The starting configuration is chosen by independent biased coin flips with probability
		$p=0.1$ of having $\symb{1}$ at each site.
		Ignoring a negligible set of time steps (represented by gray lines),
		the configuration quickly becomes almost uniform.
	}
	\label{fig:xor:randomization}
	\end{center}
\end{figure}

Yet, the most remarkable property of the XOR cellular automaton for us
is its \emph{randomization effect}:
if $x$ is chosen using independent flips of a \emph{biased} coin
(say, with probability $p\in (0,1)$ of having $\symb{1}$ at each site),
then $T$ will gradually \emph{randomize} $x$,
meaning $T^t x$ will be asymptotically indistinguishable (in distribution) from
a configuration chosen using independent flips of an \emph{unbiased} coin as $t\to\infty$,
provided we ignore a negligible set of time steps $t$ (Figure~\ref{fig:xor:randomization}).

To state this more precisely, we need some notation and terminology.
A (Borel) probability measure on $\{\symb{0},\symb{1}\}^\ZZ$
is uniquely identified by the probabilities it associates to the \emph{cylinder} sets
\begin{align*}
	[a_k a_{k+1}\cdots a_{k+n}] &\isdef
		\{x\in \{\symb{0},\symb{1}\}^\ZZ: x_k x_{k+1}\cdots x_{k+n}=a_k a_{k+1}\cdots a_{k+n}\} \;.
\end{align*}
For instance, for the \emph{Bernoulli measure} with parameter $p$
(the distribution of independent flips of a biased coin with probability $p$ of having $\symb{1}$),
which we will denote by $\mu_p$,
we have
\begin{align*}
	\mu_p([a_k a_{k+1}\cdots a_{k+n}]) &=
		p^{\#_{\symb{1}}(a)} (1-p)^{\#_{\symb{0}}(a)}
\end{align*}
for any block $a=a_k a_{k+1}\cdots a_{k+n}$,
where $\#_{\symb{1}}(a)$ and $\#_{\symb{0}}(a)$ denote, respectively,
the number of $\symb{1}$s and $\symb{0}$s appearing in $a$.
The image of a probability measure $\pi$ under $T$
is another probability measure $T\pi$
with $(T\pi)(E)=\pi(T^{-1}E)$ for any measurable set $E$.
This is the distribution of $Tx$ if $x$ is chosen at random according to $\pi$.
A sequence of probability measures $\nu_1,\nu_2,\ldots$
is said to \emph{converge weakly} to a measure $\pi$
if the probabilities that $\nu_t$ associate to each fixed cylinder
converge to the probability of that cylinder according to $\pi$.

By the above-mentioned balance property, $T\mu_{1/2}=\mu_{1/2}$.
Miyamoto~\cite{Miy79} and Lind~\cite{Lin84} (following experimental observations
made by Wolfram~\cite{Wol83})
proved that
\begin{theorem}
\label{thm:xor:randomization}
	There is a set $J\subseteq\NN$ of density~$1$ such that
	for every $p\in (0,1)$, $T^t\mu_p \to \mu_{1/2}$
	as $t\to\infty$ within $J$.
\end{theorem}
Here, the \emph{density} of a set of non-negative integers $J$
is defined as 
\begin{align*}
	d(J) &\isdef \lim_{n\to\infty}\frac{\abs{J\cap [0,n)}}{n}
\end{align*}
when the limit exists.
The theorem in particular implies that
the Ces\`aro averages $\frac{1}{n}\sum_{t=0}^{n-1} T^t\mu_p$
converge to $\mu_{1/2}$ as $n\to\infty$.

The randomization behaviour of the XOR cellular automaton can be seen
as an analogue (or an instance) of the second law of thermodynamics: 
\emph{the system evolves towards an equilibrium
in a macroscopic state with highest degree of disorder}.
Here, the term \emph{macroscopic} is understood as synonymous with \emph{statistical}:
the \emph{macroscopic state} of a configuration $x$ consists of the frequency
of occurrence of every finite word $a\in\{\symb{0},\symb{1}\}^*$ in $x$.
This information is encapsulated conveniently in a translation-invariant probability measure $\pi_x$
that is defined by those frequencies and which has $x$ as a ``typical element''.
The \emph{equilibrium state} (the uniform Bernoulli measure)
is the least presumptive (most random) state: every word of length $n$
has the same frequency $2^{-n}$.
In Sections~\ref{sec:macro-states} and~\ref{sec:disordered-states},
we shall make this interpretation more precise.

The starting configuration does not need to be Bernoulli
for the XOR cellular automaton to randomize it.
A random configuration which is a realization of
a (bi-infinite) $k$-step Markov chain with positive transition probabilities
is also randomized by the XOR cellular automaton.
In other words, the conclusion of Theorem~\ref{thm:xor:randomization}
remains true if $\mu_p$ is replaced with a full-support Markov measure~\cite{FerMaaMarNey00}.
More generally, randomizaton is known to occur
as long as the starting measure is \emph{harmonically mixing}~\cite{PivYas02,PivYas04}.\footnote{%
	For the definition of harmonic mixing and basic properties of
	the class of harmonically mixing measures, see~\cite{PivYas02,PivYas04}.
	The result of~\cite{FerMaaMarNey00} covers also the measures
	with \emph{complete connections} and \emph{summable decay}.
	These, however, turn out to be included in the class of harmonically mixing measures~\cite{HosMaaMar03}.
}
A complete characterization of the measures randomized by the XOR map is nevertheless missing.

\subsection{A reversible cellular automaton}
\label{sec:examples:reversible}

\begin{figure}
	\begin{center}
	\includegraphics[scale=0.04]{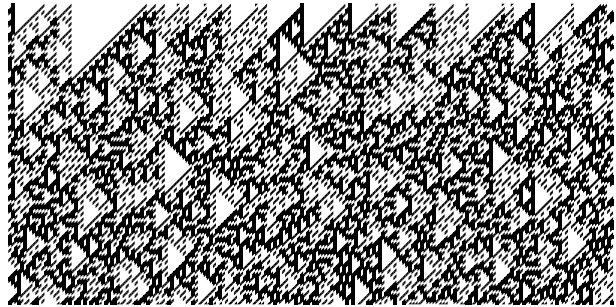}
	\caption{
		Randomization effect of the reversible cellular automaton of Maass and Mart\'inez.
		In the initial configuration, each component of the symbol at each site
		is chosen with an independent coin flip with probability $0.1$ of having a $\symb{1}$.
	}
	\label{fig:reversible-additive}
	\end{center}
\end{figure}

The analogy with the second law of thermodynamics would have been stronger
if the XOR cellular automaton were reversible\footnote{%
	A cellular automaton is said to be \emph{reversible} if
	it is bijective and has another cellular automaton as inverse.
	This is equivalent to bijectivity, because the configuration space is compact and metric.
} or symmetric under time reversal.\footnote{%
	A reasonable definition of \emph{time-reversal symmetry} for cellular automata is that
	$T$ is reversible and there is another
	reversible cellular automaton $R$ such that $T^{-1}=R^{-1} T R$;
	see~\cite{GajKarMor12}.
}
Consider now a different cellular automaton acting on the configurations
of symbols from $S\isdef\{\symb{0},\symb{1}\}\times\{\symb{0},\symb{1}\}$.
Each site $i\in\ZZ$ of a configuration $(x,y)$ carries two symbols $x_i$ and $y_i$,
and the cellular automaton map $T$ is defined by
$\left(T(x,y)\right)_i \isdef (y_i,x_i+y_{i+1})$, where the addition is again modulo~$2$.\footnote{%
	A reader familiar with Kac's ring model (see~\cite{Kac59}, Section~III.14) 
	might recognize some similarity.
	The infinite version of Kac's model
	can be defined with $\left(T(x,y)\right)_i \isdef (x_i,x_{i+1}+y_{i+1})$.
	The first component represents the presence or absence of marks on each site
	and the second the colour of the balls.
}
Let us call this the \emph{XOR-transpose} cellular automaton.\footnote{%
	The name is suggested by the fact that the space-time diagrams of this cellular automaton
	are obtained from the space-time diagrams of
	a variant of the XOR cellular automaton with $(Tx)_i\isdef x_{i-1}+x_{i+i} \pmod{2}$
	by exchanging the role of time and space.
}
As in the XOR example, the map $T$ is continuous and translation-invariant.\footnote{%
	A \emph{cellular automaton} may in fact be defined as a map on
	a symbolic configuration space $S^{\ZZ^d}$
	that is continuous and invariant under translations.
	These are precisely the maps defined homogeneously using local update rules~\cite{Hed69}.
}
It is also additive and has the balance property:
the uniform Bernoulli measure $\mu$ on $S^\ZZ$ is invariant under $T$.\footnote{%
	The balance property is shared among all cellular automaton maps that are onto~\cite{Hed69}.
}
Unlike the XOR cellular automaton,
the XOR-transpose is reversible and time-reversal symmetric:
one can traverse backwards in time 
by switching the two symbols at each site.\footnote{%
	More specifically, $T$ has an inverse given by $\left(T^{-1}(x,y)\right)_i = (y_i+x_{i+1},x_i)$.
	Setting $R(x,y)\isdef (y,x)$, we can write the inverse map $T^{-1}$
	as $R^{-1} T R$.
}
Maass and Mart\'inez~\cite{MaaMar99} proved that
the XOR-transpose cellular automaton has a similar randomization property as the XOR cellular automaton
(Figure~\ref{fig:reversible-additive}):
\begin{theorem}
\label{thm:reversible-additive:randomization}
	Let $\pi$ be the distribution of a single-step Markov chain on $S$
	with positive transition probabilities.
	Then, $\frac{1}{n}\sum_{t=0}^{n-1} T^t\pi$ converges to the uniform Bernoulli measure $\mu$
	as $n\to\infty$.
\end{theorem}
The convergence of the Ces\`aro averages 
implies the existence of a set $J\subseteq\NN$
of density~$1$ of time steps $t$ along which $T^t\pi$ converges to $\mu$
(see~\cite{JohRud95}, Corollary~1.4),
but the set $J$ might, in principle, depend on the measure $\pi$.

\subsection{A bi-permutative cellular automaton}
\label{sec:examples:bipermutative}

As a third example, let us look at the cellular automaton with symbol set
$S\isdef\{\symb{0},\symb{1},\symb{2}\}$, defined by
\begin{align}
\label{eq:permutative}
	(Tx)_i \isdef \varphi(x_{i-1},x_i,x_{i+1}) &\isdef
		\begin{cases}
			x_{i-1} + x_{i+1} + 1 \pmod{3}\quad		& \text{if $x_i=\symb{2}$,}\\
			x_{i-1} + x_{i+1} \pmod{3}				& \text{otherwise.}
		\end{cases}
\end{align}
Unlike the last two examples, the map $T$ is not additive.
Nevertheless, the local rule $\varphi$ 
is \emph{bi-permutative}, which is to say
both $a\mapsto\varphi(a,b,c)$ and $c\mapsto\varphi(a,b,c)$ are permutations.\footnote{%
	Notice that the XOR cellular automaton is also bi-permutative.
}
It follows, similarly as in the case of the XOR cellular automaton, that the map $T$ is $9$-to-$1$.
Like the last two examples, the uniform Bernoulli measure
(i.e., the distribution of a configuration chosen at random by
flipping an unbiased ``$3$-sided coin''\footnote{%
	or rolling a $3$-sided die, if you wish,
} independently for each site)
is invariant under $T$.
The bi-permutativity also ensures other statistical regularities for $T$,
similar to those enjoyed by the XOR cellular automaton~\cite{She92}.

\begin{figure}
	\begin{center}
		\includegraphics[scale=0.8]{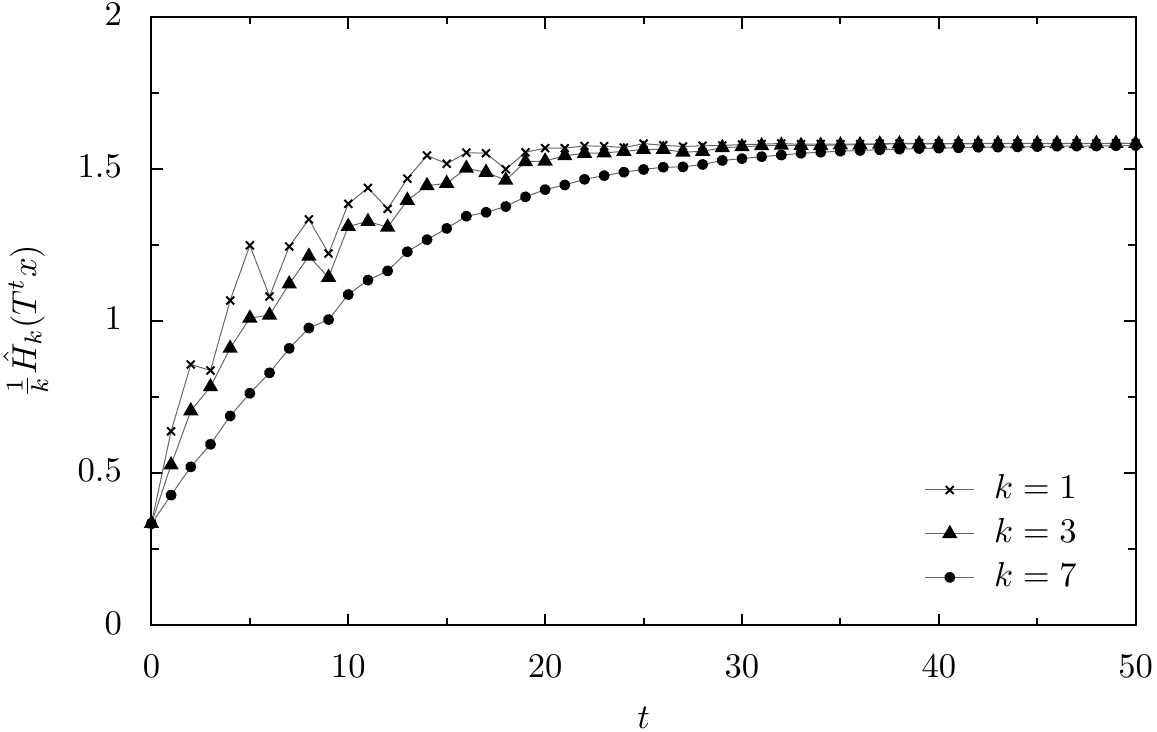}
	\caption{
		Evidence of randomization in the bi-permutative cellular automaton
		defined in Equation~(\ref{eq:permutative}).
		The starting configuration $x$ (on a ring of length $50000$)
		is chosen with independent flips of a biased $3$-sided coin with distribution
		$(\symb{0}\mapsto 0.95, \symb{1}\mapsto 0.025,\symb{2}\mapsto 0.025)$.
	}
	\label{fig:permutative:randomization}
	\end{center}
\end{figure}

It is not known whether the above cellular automaton has
a randomizing behaviour in the sense of Theorems~\ref{thm:xor:randomization}
or~\ref{thm:reversible-additive:randomization}.
Nevertheless, there is experimental evidence suggesting that
$T$ indeed randomizes biased Bernoulli configurations (Figure~\ref{fig:permutative:randomization}).
The graphs in Figure~\ref{fig:permutative:randomization} depict
the change in time of the empirical entropies of words of length $1$, $3$ and $7$
in consecutive configurations of this cellular automaton,
starting from a biased Bernoulli configuration.
More specifically, a single pseudo-random configuration $x$ is picked
by simulating independent biased ($3$-sided) coin flips,
and iterations of $T$ on $x$ are obtained for up to $50$ time steps.\footnote{%
	Instead of infinite configurations, configurations of symbols
	on a large ring (indexed by $\ZZ_N$ for $N$ large) are used.
}
At each time step, the empirical entropy of words
of length $k$ (for $k=1,3,7$) appearing in the current configuration are calculated
as follows.
For each word $w$ of length $k$ with symbols from $S$,
let $\overline{\zeta_w}(y)$ denote the frequency of appearance of
the word $w$ in configuration $y$.
The \emph{empirical entropy} of words of length $k$ appearing in $y$
is defined as
\begin{align*}
	\hat{H}_k(y) &\isdef
		-\sum_{w\in S^k} \overline{\zeta_w}(y)\log \overline{\zeta_w}(y) \;.
\end{align*}
Figure~\ref{fig:permutative:randomization} shows that
the empirical entropy $\hat{H}_k(T^t x)$ 
rapidly increases to reach its maximum at around $k\log 3$,
where it stays.

The empirical entropy $\hat{H}_k(y)$ is a measure of disorder in $y$.
It is maximized when all words of length $k$ appear in $y$ with
approximately the same frequency.  For instance, a configuration chosen using
independent unbiased coin flips (which is considered to be maximally disordered)
has, with probability~$1$,
the maximum empirical entropy $\hat{H}_k$ for each $k$.
The empirical entropy $\hat{H}_k(y)$ should be compared with Boltzmann's entropy
(see below).\footnote{
	For the interpretations of entropy,
	see e.g.~\cite{CovTho91,Geo03}.
}
Although not exhaustive,
the simulation in Figure~\ref{fig:permutative:randomization}
suggests a gradual approach towards a maximally disordered state.

\subsection{Rule~30}
\label{sec:examples:rule30}

Yet another example where randomization seems to be present is
the so-called \emph{Rule~30} cellular automaton.  The Rule~30 cellular automaton
has the binary alphabet $\{\symb{0},\symb{1}\}$ as the symbol set
and may be defined by the logical expression
\begin{align*}
	(Tx)_i\isdef \varphi(x_{i-1},x_i,x_{i+1}) &\isdef x_{i-1} \XOR (x_i \OR x_{i+1}) \;,
\end{align*}
where $\XOR$ denotes ``exclusive or''.
The local rule $\varphi$ 
is not bi-permutative, but it is \emph{left-permutative}
(i.e., $a\leftrightarrow\varphi(a,b,c)$ is a bijection for each $b$ and $c$).
This still implies that the map $T$ is onto, and that
each configuration has at most $4$ pre-images under $T$.
It follows that the Rule~30 cellular automaton again has the balance property.
Starting from an unbiased Bernoulli configuration,
the Rule~30 cellular automaton enjoys similar statistical regularities
as in the previous examples, along almost all space-time directions~\cite{She97}.\footnote{%
	More specifically, $T\sigma^k$ is an exact endomorphism unless $k=-1$.
}

This cellular automaton was first studied by Wolfram~\cite{Wol86}.
He noticed that even with a simple starting configuration,
the iterations of the Rule~30 cellular automaton produce
complex seemingly unpredictable patterns.
He proposed a method for generation of pseudo-random sequences
by initializing the Rule~30 cellular automaton with a ``seed''
and picking the symbols appearing on a particular site every few time steps,
which he tested against standard statistical randomness tests.\footnote{%
	Rule~30 is in fact used in Mathematica as one of the methods
	for pseudo-random number generation.
}

\begin{figure}
	\begin{center}
		\includegraphics[scale=0.8]{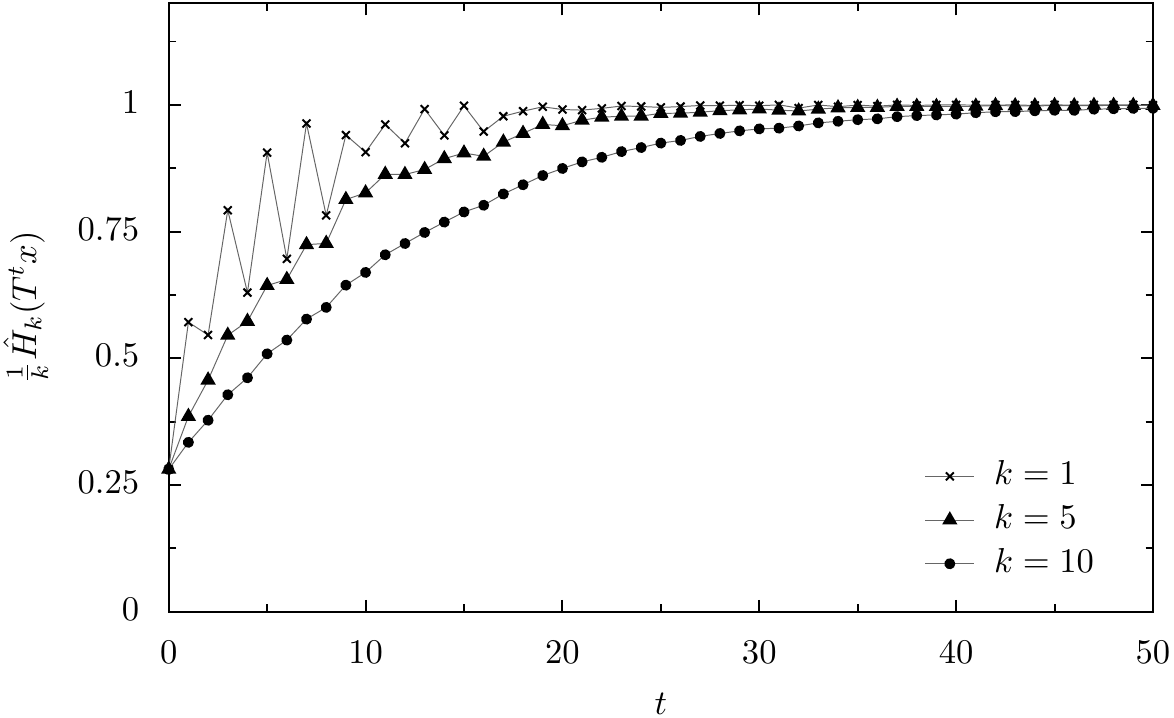}
	\caption{
		Evidence of randomization in the Rule~30 cellular automaton.
		The starting configuration $x$ (on a ring of length $50000$)
		is chosen with independent flips of a biased coin with probability $p=0.05$
		of having a $\symb{1}$.
	}
	\label{fig:rule30:randomization}
	\end{center}
\end{figure}

Figure~\ref{fig:rule30:randomization} shows evidence
for randomization in the Rule~30 cellular automaton
starting from biased Bernoulli configurations.
The empirical entropies are calculated as in the previous example.

\subsection{Q2R spin dynamics}
\label{sec:examples:Q2R}

One feature that is common among the first three examples
(and is suspected for the Rule~30 cellular automaton)
is the absence of conserved energy-like quantities~\cite{ForKarTaa11}.
A non-trivial conserved quantity would partition
the macroscopic states into unescapable fibers,
hence preventing complete randomization.
Nevertheless, one might still expect randomization
within each fiber.

The next example is based on the configurations of the Ising model,
and was introduced by Vichniac~\cite{Vic84}.
The \emph{Ising model} is a model of ferromagnetism:
each site of an infinite two-dimensional square lattice (indexed by $\ZZ^2$)
carries a symbol $\upspin$ or $\downspin$,
representing two possible directions of a \emph{magnetic spin}.
The interaction between spins is modelled by assigning
energy $-1$ or $+1$ to any pair of neighbouring sites
whose symbols are, respectively, aligned or anti-aligned spins.
The energy content of a region is the sum of the interaction energy
of neighbouring sites in that region.
Hence, lower energy in a region corresponds to
an average tendency of neighbouring spins to be aligned.

The dynamics is through alternate applications of two maps $T_0$ and $T_1$:
the first map updates the \emph{even} sites (i.e., the sites $(i,j)$ with $i+j$ even)
and the second updates the \emph{odd} sites.
The updating is done in such a way that the energy is preserved:
a spin is flipped if and only if the flipping does not change the total energy
of the site and its four immediate neighbours.  More specifically,
let us say that a site $(i,j)$ is \emph{balanced} on a configuration $x$
if half of the neighbouring spins $x_{i+1,j}$, $x_{i,j+1}$, $x_{i-1,j}$ and $x_{i,j-1}$
are upward and the other half are downward.
For a spin $a\in\{\upspin,\downspin\}$, let $\overline{a}$ denote the spin
with the opposite direction as $a$.
Then,
\begin{align*}
	(T_0 x)_{i,j} &\isdef
		\begin{cases}
			\overline{x_{i,j}}\quad		& \text{if $i+j$ even and $(i,j)$ balanced on $x$,}\\
			x_{i,j}						& \text{otherwise,}
		\end{cases}
\end{align*}
and similarly for $T_1$.
The dynamical system defined by the composition $T\isdef T_1T_0$
is called the \emph{Q2R} cellular automaton.\footnote{%
	Strictly speaking, this is not a cellular automaton
	with the common definition of the term, because the
	even and odd sites are treated differently.
	It can however be recoded into a standard cellular automaton.
}

The Q2R system is reversible and symmetric under time reversal
in the sense that $T^{-1}=T_0 T_1$.
By construction, it also conserves the energy.
The \emph{conservation of energy} can be formulated in various equivalent ways.
For us, it suffices to say that $T$ (indeed, each of $T_0$ and $T_1$)
keeps the \emph{average energy per site} 
invariant.  Note that the average energy per site of a configuration $x$
is a function of its macroscopic state $\pi_x$.
The set of translation-invariant probability measures with
a given average energy per site is convex and closed under the topology of
weak convergence.  Therefore, any limit or Ces\'aro limit of the measures $T^t\pi_x$
will have the same average energy per site as $\pi_x$.

As before, we consider the uniform Bernoulli measure on
the configuration space $\{\upspin,\downspin\}^{\ZZ^2}$
to be a representation of a ``maximally disordered'' state, because it assigns
the same probability $2^{-\abs{A}}$ to all cylinder sets
\begin{align*}
	[q_A] &\isdef
		\{x\in \{\upspin,\downspin\}^{\ZZ^2}: \text{$x_i=q_i$ for $i\in A$}\} \;.
\end{align*}
Put another way, in a typical spin configuration obtained by independent unbiased coin flips,
(translations of) each finite pattern $q_A:A\to\{\upspin,\downspin\}$
appears with the same frequency $2^{-\abs{A}}$.
Another way to express this is to say that the \emph{entropy}
\begin{align*}
	H_A(\mu) &\isdef
		-\sum_{q_A:A\to\{\upspin,\downspin\}} \mu([q_A])\log \mu([q_A])
\end{align*}
of any \emph{finite window} $A\subseteq\ZZ^2$ has its maximum value $\abs{A}\log 2$
if $\mu$ is the uniform Bernoulli measure.

The description of a ``maximally disordered'' state with a given average energy per site
is more subtle.  Indeed, since the constraint is not local, it might not be possible
to maximize the entropy $H_A(\mu)$ simultaneously for all finite windows $A$.
However, if $B\supseteq A$, a larger value for $H_B(\mu)$ is a better indication of disorder
than a larger value for $H_A(\mu)$.
Therefore, one may measure the disorder
by the limit \emph{entropy per site}
\begin{align*}
	h(\mu) &\isdef
		\lim_{n\to\infty} \frac{H_{I_n}(\mu)}{\abs{I_n}} \;,
\end{align*}
where $I_n=[-n,n]\times[-n,n]$.\footnote{%
	For translation-invariant measures,
	the limit exists and is equal to $\inf_{n} \frac{1}{\abs{I_n}}H_{I_n}(\mu)$.
}
A maximally disordered state with a given average energy per site
may therefore be identified with an ergodic translation-invariant measure that
has the prescribed average energy per site and maximal entropy per site
subject to the energy constraint.
These are the ergodic \emph{Gibbs measures}
for the Ising model (see e.g.~\cite{Rue04,Isr79}).\footnote{%
	In the standard Ising model, the ergodic Gibbs measures are considered to be
	suitable descriptions of the macroscopic states of equilibrium
	for the ferromagnetic material (see e.g.~\cite{Geo88}).  
}

\begin{figure}
	\begin{center}
		\begin{tabular}{ccccc}
			\begin{minipage}[c]{0.31\textwidth}
				\centering
				\includegraphics[scale=0.25]{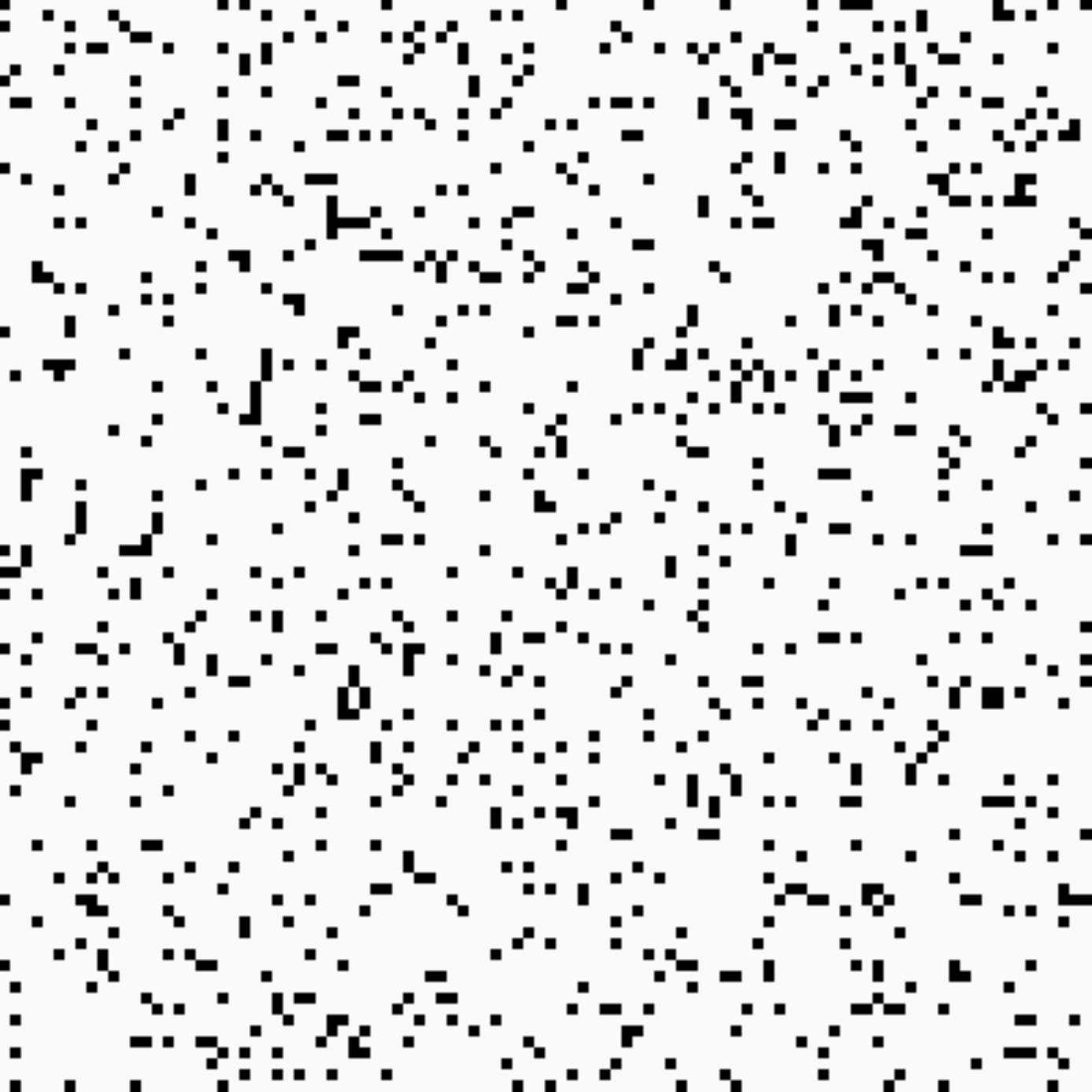} 
			\end{minipage} & &
			\begin{minipage}[c]{0.31\textwidth}
				\centering
				\includegraphics[scale=0.25]{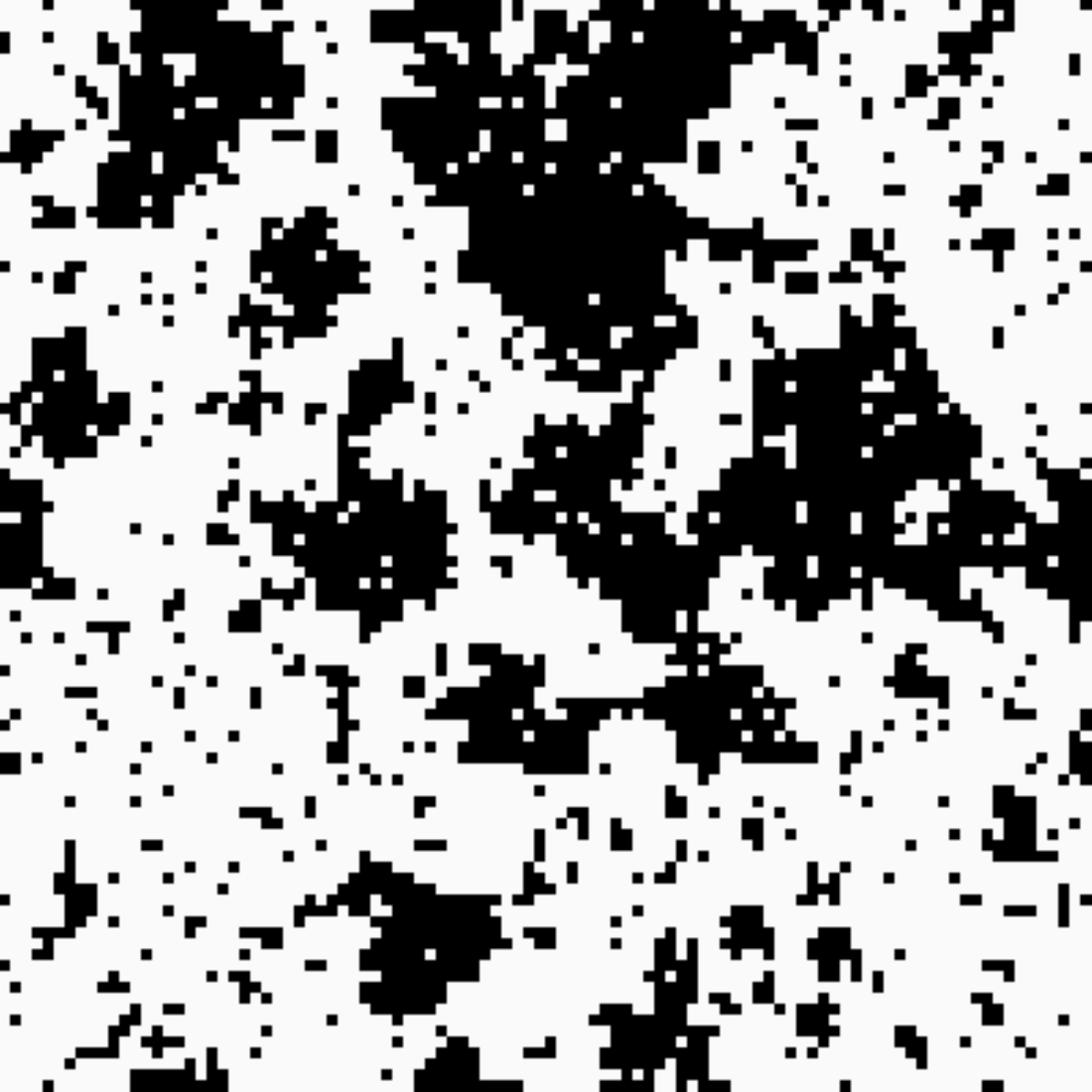}
			\end{minipage} & &
			\begin{minipage}[c]{0.31\textwidth}
				\centering
				\includegraphics[scale=0.25]{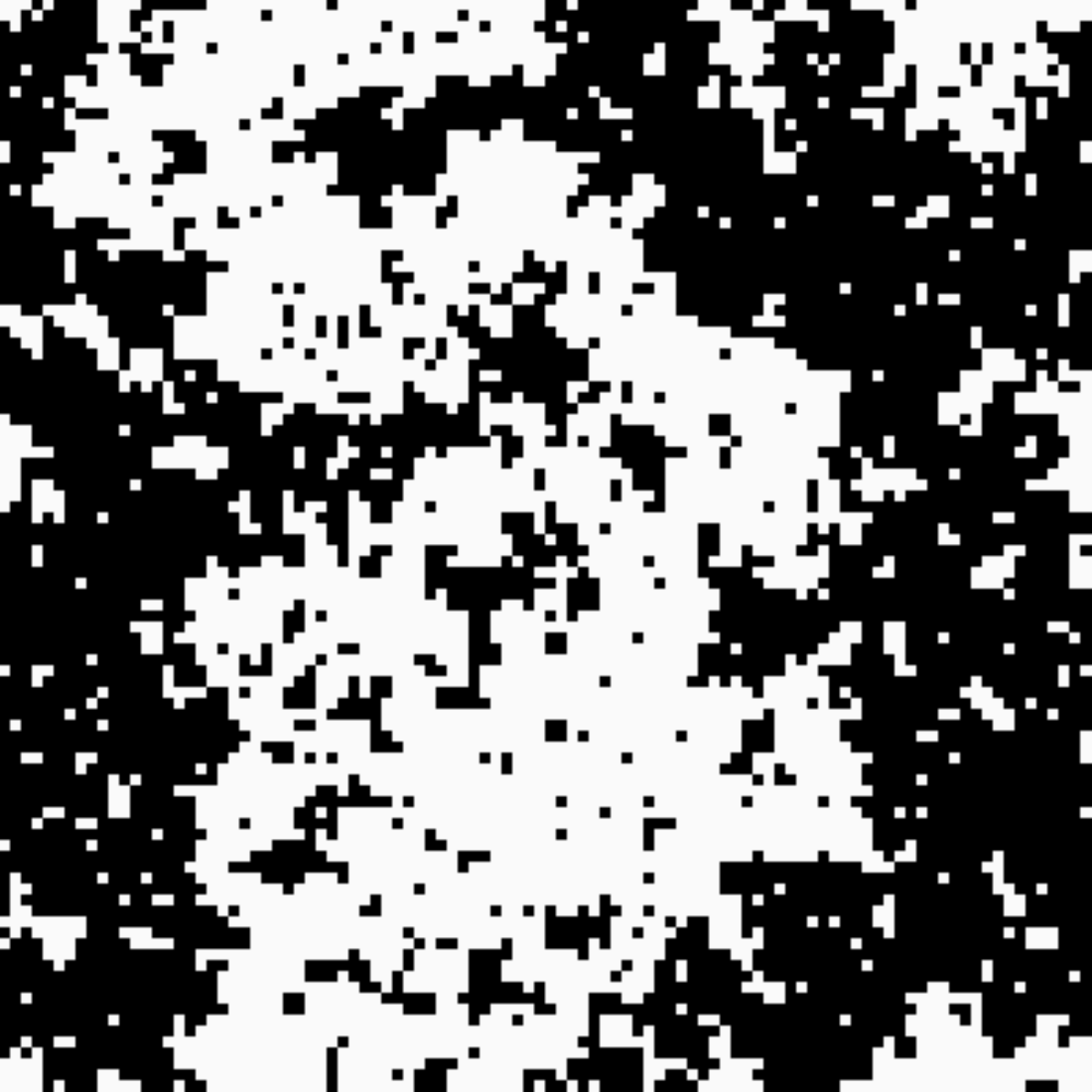}
			\end{minipage} \medskip\\
			t=0 && t=500 && t=5000
		\end{tabular}	
	\caption{
		Simulation of the Q2R cellular automaton
		starting from a coin-flip configuration with probability $p=0.1$
		of having $\upspin$ (represented by black).
		After a relatively short while, the macroscopic look of the configuration
		seems to reach an equilibrium with upward and downward spins clustered together.
	}
	\label{fig:Q2R:simulation}
	\end{center}
\end{figure}

Figure~\ref{fig:Q2R:simulation} shows few snapshots from a simulation of the Q2R
cellular automaton starting from a biased Bernoulli configuration.
At the beginning, the spins gradually cluster, even though the total length of
the boundaries between upward and downward clusters remains constant.
After a while, the macroscopic picture of the configuration appears to reach
an equilibrium, which resembles a typical configuration chosen according
to a Gibbs measure of the Ising model.\footnote{%
	Again, the simulation is made on a torus $\ZZ_N^2$ rather than the infinite lattice $\ZZ^2$.
	The ``equilibrium'' configurations could be compared with a random configuration
	generated by a Gibbs sampler for the Ising model.
}
More elaborate simulations have shown numerical agreement
with the Ising model (see e.g.~\cite{Her86,TofMar87}), hence supporting the conjecture
that Q2R indeed randomizes a coin-flip configurations within the corresponding
average energy per site level.

\bigskip

In the next two sections, we attempt to make the concepts of macroscopic state
and maximally disordered state more precise.

\section{Macroscopic States}
\label{sec:macro-states}

Let us fix a symbol set $S$ and denote by $\mathcal{X}=\{x:\ZZ^d\to S\}$
the set of all $d$-dimensional configurations of symbols from $S$.
A configuration $x\in\mathcal{X}$ is considered as a \emph{microscopic state} of
a system.
The \emph{macroscopic state} of $x$ consists of all information in $x$
that is visible through ``macroscopic observations''.
Which observations are considered macroscopic is somewhat arbitrary
and depends on the physical context.  Here, we equate ``macroscopic'' with ``statistical'':
a macroscopic observation would amount to identifying the frequency of
a fixed pattern, or the spatial average of a ``microscopic observation''.

To be more specific, let us call a function $f:\mathcal{X}\to\RR$
a \emph{local observable} if the symbols $x_i$ at finitely many sites $i\in A$
are sufficient to determine $f(x)$.  For instance, if $q:A\to S$
is a pattern on a finite set $A$, the function $x\mapsto\zeta_q(x)$ that has value $1$
if $x$ agrees with $q$ on $A$ and $0$ otherwise is a local observable.
Furthermore, any local observable is a linear combination of observables of this type.

If $f$ is a local observable, the spatial average
\begin{align}
\label{eq:spatial-average}
	\overline{f}(x) &\isdef
		\lim_{n\to\infty} \frac{1}{\abs{I_n}}\sum_{i\in I_n} f(\sigma^i x)
\end{align}
will be considered as a \emph{macroscopic observable}.
As before, $I_n\isdef[-n,n]^d$, and $\sigma^i$ denotes the translation by $i$.
For instance, $\overline{\zeta_q}(x)$ is simply the frequency of
the occurrence of (translations of) the pattern $q$ in $x$.  
The limit in~(\ref{eq:spatial-average}) may or may not exist.
If well-defined, the collection $\left(\overline{f}(x): \text{$f$ local}\right)$
defines a unique translation-invariant probability measure $\pi_x$ with
\begin{align*}
	\pi_x(f) = \int f\xd\pi_x &= \overline{f}(x) \;,
\end{align*}
describing the statistics of $x$.
In particular, $\pi_x([q])=\overline{\zeta_q}(x)$ for any finite pattern $q$.

Every translation-invariant probability measure on $\mathcal{X}$
arises from a configuration in the above fashion~\cite{Oxt63}.
Nevertheless, not every translation-invariant probability measure
should be considered as an unambiguous macroscopic state.
To illustrate this, consider a one-dimensional configuration $z$
with $z_i=\symb{0}$ for $i<0$ and $z_i=\symb{1}$ for $i>0$.
Then $\pi_z=\frac{1}{2}(\delta_{\underline{\symb{0}}}+\delta_{\underline{\symb{1}}})$,
where $\delta_{\underline{\symb{0}}}$ and $\delta_{\underline{\symb{1}}}$
are the point-mass measures at the uniform configurations
$\underline{\symb{0}}$ and $\underline{\symb{1}}$.
The measure $\frac{1}{2}(\delta_{\underline{\symb{0}}}+\delta_{\underline{\symb{1}}})$
however suggests an ambiguous situation in which we are uncertain
about which of $\underline{\symb{0}}$ and $\underline{\symb{1}}$
is the real configuration of the system.\footnote{%
	See~\cite{Geo88}, Paragraph~(7.8),
	for a similar reasoning.
}
The ambiguity comes from the fact that the configuration $z$
lacks homogeneity: its left and right tails have different macroscopic looks.\footnote{%
	As an example in which inhomogeneity does not
	arise from left-right asymmetry,
	let $m_0=0$ and $m_k\isdef 2(1 + 2^2 + \cdots + k^2)$,
	and construct a one-dimensional configuration $z:\ZZ\to\{\symb{0},\symb{1}\}$ with
	$z_i=\symb{0}$ if $m_k\leq\abs{i}< m_k + (k+1)^2$ and
	$z_i=\symb{1}$ if $m_k + (k+1)^2 \leq\abs{i}< m_{k+1}$.
	Then, again
	$\pi_z=\frac{1}{2}(\delta_{\underline{\symb{0}}}+\delta_{\underline{\symb{1}}})$.
}

Here, we focus on configurations that are homogeneous.
We call a configuration~$x$ \emph{homogeneous}\footnote{%
	Such points are called \emph{regular} in~\cite{Oxt52}.
} if
\begin{enumerate}[i)]
	\item $\pi_x$ is well-defined (i.e., the spatial average $\overline{f}(x)$
		of every local observable $f$ exists on $x$),
	\item $\pi_x$ cannot be written as a non-trivial convex combination of other
		translation-invariant measures (i.e., $\pi_x$ is ergodic for the group of translations),\footnote{%
			It might not be intuitively clear why $\pi_x$ should be required to be ergodic
			in order for $x$ to be called homogeneous.
			A perhaps more plausible condition equivalent to the ergodicity of $\pi_x$ is that
			for every local observable $f$ and each $\varepsilon>0$,
			the upper density of the set
			\begin{align*}
				\left\{a\in\ZZ^d:
					\abs{\frac{1}{\abs{I_m}}\sum_{i\in a+I_m} f(\sigma^i x) - \overline{f}(x)}
					> \varepsilon
				\right\}
			\end{align*}
			in $\ZZ^d$ goes to $0$ as $m\to\infty$~\cite{Oxt52}.
			Note that for both examples of inhomogeneous configurations mentioned above,
			this condition fails for the function
			$f=\zeta_{\symb{1}}$ with $f(x)=1$ if $x_0=\symb{1}$ and $f(x)=0$ otherwise.
		} and
	\item $x$ is a \emph{point of density} for $\pi_x$, which is to say,
		every finite pattern occurring in $x$ occurs with positive frequency.
\end{enumerate}
The measure $\pi_x$ describing the statistical averages of
a homogeneous configuration $x$ will be called the \emph{macroscopic state} of $x$.
The set of homogeneous configurations sharing the same ergodic measure $\pi$
as macroscopic state is called the \emph{ergodic set} of $\pi$.
The countability of the set of finite patterns together with the ergodic theorem
implies that the ergodic set of any ergodic measure $\pi$
has measure~$1$ with respect to $\pi$ (see~\cite{Oxt52}).\footnote{%
	In particular, the set of homogeneous configurations has measure~$1$
	with respect to any translation-invariant probability measure.
}
Hence, one may think of a configuration in the ergodic set of $\pi$
as a \emph{typical} configuration with macroscopic state $\pi$.\footnote{%
	If need be,
	stronger notions of homogeneity and typicalness can be obtained
	by intersecting the ergodic set of $\pi$ with other sets of measure~$1$.
}

\section{Maximally Disordered States}
\label{sec:disordered-states}

From the definition, it follows that, for any finite window $A\subseteq\ZZ^d$,
the entropy $H_A(\pi)$ of a macroscopic state (i.e., a translation-ergodic measure) $\pi$
agrees with the empirical entropy $\hat{H}_A(x)$ of any configuration $x$
that is typical for $\pi$.
The entropy $H_A(\pi)$ is a convex continuous function of $\pi$,
taking its maximum value $\abs{A}\log\abs{S}$ only when $\pi$ assigns equal probabilities
to every cylinder with base $A$.

The limit entropy per site $h(\pi)$ is affine (hence convex)
and takes its maximum value $\log\abs{S}$ precisely when $\pi$ is the uniform Bernoulli measure,
that is the state with ``maximum disorder''.
The map $\pi\mapsto h(\pi)$ is however not continuous.
For example, for each $m>0$, let $x(m)$ be a periodic configuration in $\{\symb{0},\symb{1}\}^\ZZ$
that has each word of length $m$ exactly once in its period.\footnote{%
	Such a configuration corresponds to an Eulerian circuit
	on the de~Bruijn graph of order~$m$.
} Then, the macroscopic states $\pi_{x(m)}$ have $0$ entropy per site
yet converge weakly to the uniform Bernoulli measure as $m\to\infty$.
In fact, every macroscopic state is a weak limit of
macroscopic states of periodic configurations (which all have entropy~$0$).
Nevertheless, entropy per site is upper semi-continuous:
$\pi_n\to\pi$ implies $\limsup_{n\to\infty}h(\pi_n)\leq h(\pi)$ (see e.g.~\cite{Wal82}).

Let us now consider a concept of energy as in the Ising model.
Namely, let $f:\mathcal{X}\to\RR$ be a local observable,
representing the energy contribution of the symbol at the origin
when interacting with the nearby symbols.  For instance,
for the Ising model, we can set
\begin{align*}
	f(x) &= \begin{cases}
		\frac{1}{2}\left(n_{\downspin}(\partial_0 x) - n_{\upspin}(\partial_0 x)\right) \quad
			& \text{if $x_0=\upspin$,} \\
		\frac{1}{2}\left(n_{\upspin}(\partial_0 x) - n_{\downspin}(\partial_0 x)\right)
			& \text{if $x_0=\downspin$,}
	\end{cases}
\end{align*}
where $n_{\upspin}(\partial_0 x)$ and $n_{\downspin}(\partial_0 x)$
are the numbers of upward and downward spins among the four
neighbours $x_{1,0}$, $x_{0,1}$, $x_{-1,0}$ and $x_{0,-1}$ of site $0$.
Then, $\overline{f}(x)$ represents the average energy per site of
a configuration $x$, which is well-defined if $x$ is homogeneous,
and agrees with $\pi_x(f)$.

Suppose $e$ is a real number within the range of $\overline{f}$.
Among all the macroscopic states $\pi$ satisfying $\pi(f)=e$,
those with maximum entropy per site $h(\pi)$ could be considered as the \emph{most disordered}.
These are the presumed \emph{equilibrium states} of a system
in which the energy $f$ is conserved.\footnote{%
	A similar discussion applies if rather than single notion of energy,
	we have a finite number of conserved quantities $f_1,f_2,\ldots,f_n$.
}
Applying the Lagrange multipliers method
(Legendre transform),
the optimization problem
\begin{align*}
	\textsf{maximize}\quad		& h(\pi)\\
	\textsf{subject to}\quad	& \pi(f)=e
\end{align*}
(for $e$ in the range of $\overline{f}$)
translates into the unconstrained problem
\begin{align}
\label{eq:var-problem}
	\textsf{maximize}\quad		& h(\pi) - \beta\pi(f)
\end{align}
(for $\beta\in\RR$).
The compactness of the space,
the continuity of $\pi\mapsto\pi(f)$ and the upper semi-continuity of $\pi\mapsto h(\pi)$
imply that, in both problems, the maximums are achieved by some
translation-invariant probability measures.\footnote{%
	However, the maximum in the first problem is not necessarily achieved
	by ergodic measures (i.e., macroscopic states).
	Such a situation corresponds to a first-order phase transition
	(see~\cite{Rue04,Isr79}).
}
Dobrushin, Lanford and Ruelle proved that
the macroscopic states solving the variational problem~(\ref{eq:var-problem})
are precisely the ergodic \emph{Gibbs measures} at \emph{inverse temperature}~$\beta$.\footnote{%
	In the current setting,
	Gibbs measures coincide with full-support Markov measures.
}
See~\cite{Rue04,Isr79,Kel98} for more information.

\section{Boltzmann's Theory and Cellular Automata}

Let us take a moment to draw parallels between the concepts in Boltzmann's gas theory
and in cellular automata.
We refer to the survey article of the Ehrenfests~\cite{EhrEhr12}
and the book by Kac~\cite{Kac59}, which contain excellent accounts
of Boltzmann's theory and related issues.
See also~\cite{Leb99} for a general discussion.

Boltzmann considered an isolated collection of $n$ particles
(identical hard spheres) interacting via elastic collisions.\footnote{%
	See Chapter~I of~\cite{EhrEhr12} and Sections~III.1--2 of~\cite{Kac59}.
}
The particles are assumed to be homogeneously distributed
in (a bounded but large region of) the space.
To be concrete, we may consider
a cubic region with periodic boundary conditions.
The focus is thus only on the velocity of the particles.
Assuming that the number of particles is very large,
we take $\rho(v,t)\xd v$ to be the fraction of particles that,
at time $t$, have velocities within an infinitesimal approximation $\xd v$
of each value $v\in\RR^3$.
Using the assumption of spatial homogeneity, Boltzmann estimated
the average number of collisions, in an infinitesimally small time interval $(t,t+\xd t)$,
among particles with velocities close to $u$ and those with velocities close to $v$
(the \emph{Stosszahlansatz}).\footnote{%
	More specifically, the \emph{Stosszahlansatz} says that
	the frequencies of particles with different velocities that enter an infinitesimally small region
	at any given time are statistically independent.
}
The model of elastic collisions (the conservation of energy and momentum)
could now be invoked to obtain the new distribution $\rho(v,t+\xd t)\xd v$
for the velocity of the particles at the end of the time interval $(t,t+\xd t)$.
This leads to an equation describing the time evolution of $\rho(v,t)$ known as the Boltzmann equation.
Boltzmann used this equation to show that the quantity
$-\int \rho(v,t)\log \rho(v,t)\xd v$ is monotonically increasing in time,
except at an equilibrium in which the velocities are distributed according to the Maxwell distribution
$\rho(v)\sim \xe^{-c \abs{v}^2}$.

Boltzmann's derivation of the ``law of increase in entropy''
faced two major criticisms.\footnote{%
	See Section~7 of~\cite{EhrEhr12} and Section~III.7 of~\cite{Kac59}.
}
Loschmidt objected that a system governed by
a reversible and time-reversal symmetric dynamics 
(like a system of particles interacting via elastic collisions)
cannot possibly have an observable that is
invariant under time reversal (like entropy) and is
monotonically increasing in all situations.
Zermelo's objection was based on Poincar\'e's recurrence theorem.
According to Liouville's theorem, a Hamiltonian system (such as a system of particles)
preserves the phase space volume (i.e., the Lebesgue measure).
Poincar\'e's theorem states that in a volume-preserving system
whose phase-space has finite volume
(such as an isolated system of particles in a $3$-dimensional torus),
almost every trajectory eventually returns (infinitely many times)
arbitrarily close to its starting point.
This again implies that such a system cannot have a \emph{continuous} observable
that is monotonically increasing in time for \emph{almost all} starting points.

In order to address these criticisms,
Boltzmann later introduced another more refined framework.\footnote{%
	See Chapter~II of~\cite{EhrEhr12} and Section~III.8 of~\cite{Kac59}.
}
In this new setting, each particle~$i$ is described by a state variable $x_i$,
which could, for instance, consist of the position as well as the velocity of the particle.
The phase space of an individual particle (i.e., the range of values of $x_i$)
is divided into small equally-sized regions $A_1,A_2,\ldots$.
Given a configuration of particles $x$,
we can form the fraction $\rho_k= n_k/n$ of particles
whose states are in region $A_k$.
Conversely, given the macroscopic information $\rho=(\rho_1,\rho_2,\ldots)$,
there corresponds a set $[\rho]$ consisting of all particle configurations
that have fractions $\rho_1,\rho_2,\ldots$ of particles in regions $A_1,A_2,\ldots$.
If the number of particles is very large,
the volume of the set $[\rho]$ could be measured by the
quantity $H(\rho)=-\sum_k \rho_k\log \rho_k$.
Neglecting any interaction energy between particles, the energy
of a configuration $x$ could be written as $E(x)=n\sum_k \rho_k e_k$,
where $e_k$ is the (approximate) energy of a particle whose state is within $A_k$.
Boltzmann now argued that a system with energy $E$ at equilibrium
is most likely to be found (at almost any moment of time)
to have a state distribution $\rho$ that maximizes the quantity $H(\rho)$
among all the state distributions with energy $E$, for this is the distribution
for which $[\rho]$ takes the overwhelmingly largest portion of
the set of all configurations with energy $E$.
If $n$ is large, this equilibrium distribution is (approximately) given by
$\rho_k\sim \xe^{-\beta e_k}$, where $\beta$ is a Lagrange multiplier
for tuning~$E$.

The analogy with cellular automata should be clear.
Rather than particles, the elementary pieces of information in cellular automata
are carried by lattice sites representing discretized positions in the space.
The symbol at site $i$ should therefore be compared with the state of particle $i$.
The model of elastic collisions governing the interaction between the particles
is replaced with the local update rule describing
the cellular automaton map $T:S^{\ZZ^d}\to S^{\ZZ^d}$.
The fraction $\overline{\zeta_a}(x)$ of sites having symbol $a$
is an elementary macroscopic observable analogous to the fraction $n_k/n$
of particles whose states are in region $A_k$, but now it is clear that
one must also take into account the macroscopic observables $\overline{\zeta_q}(x)$
(for larger patterns $q:A\to S$)
which contain information about correlations between finite collections of sites.
Boltzmann's entropy corresponds to the empirical entropy $\hat{H}_0(x)$
of symbols appearing in the configuration $x$, or more generally,
the empirical entropy $\hat{H}_A(x)$,
which measures lack of bias in the frequency of the patterns with support $A$ occurring in $x$.

To understand Boltzmann's argument about the increase in entropy,
consider the XOR cellular automaton (Section~\ref{sec:examples:xor}),
and let $x$ be a Bernoulli random configuration
with parameter $p\in(0,1)$
(i.e., a homogeneous configuration whose macroscopic state is
described by the Bernoulli measure with parameter $p$).
In particular, the words of length $2$ have frequencies
\begin{align*}
	\overline{\zeta_{\symb{0}\symb{0}}}(x) &= (1-p)^2 \;, &
	\overline{\zeta_{\symb{0}\symb{1}}}(x) &= (1-p)p \;, &
	\overline{\zeta_{\symb{1}\symb{0}}}(x) &= p(1-p) \;, &
	\overline{\zeta_{\symb{1}\symb{1}}}(x) &= p^2
\end{align*}
in $x$.
It follows that the frequency $\overline{\zeta_{\symb{1}}}(Tx)$ of occurrence
of symbol~$\symb{1}$ after one step is $\varphi(p)\isdef 2p(1-p)$.
If $\mathsf{H}(p)\isdef -p\log p - (1-p)\log (1-p)$ denotes the binary entropy function,
one can easily verify that $\mathsf{H}(\varphi(p))\geq \mathsf{H}(p)$
with equality if and only if $p=\frac{1}{2}$.
Therefore, it is indeed the case that the entropy $\hat{H}_0(Tx)$
is larger than $\hat{H}_0(x)$ unless $p=\frac{1}{2}$.\footnote{%
	Similar (but more cumbersome) calculations lead to the same conclusion for the examples
	in Sections~\ref{sec:examples:bipermutative} and~\ref{sec:examples:rule30}.
	More generally, one can show that if $f:S\times S\to S$
	is a function that is permutative on both its arguments,
	$X$ and $Y$ are independent $S$-valued random variables with distributions $p$ and $p'$,
	and $Z=f(X,Y)$, then $H(Z)\geq H(X)$ with equality if and only if $p$ is uniform.
}

Boltzmann's assumption about the number of collisions (the \emph{Stosszahlansatz})
is analogous to the (invalid) assumption that the configuration $Tx$
is also Bernoulli, so that the frequency of occurrence
of a word $w=a_1a_2\cdots a_n$ in $Tx$ is simply the product of the frequencies
of $a_1,a_2,\ldots,a_n$.
If true, this would lead to the conclusion that the entropy
increases monotonically in the consecutive steps,
that is, $H_0(x)\leq H_0(Tx) \leq H_0(T^2x) \leq \cdots$
with the equalities only if $H_0(x)=\log 2$.
The assumption is of course false.\footnote{%
	For the XOR cellular automaton, $Tx$ is Bernoulli only if $p=\frac{1}{2}$.
}
Nevertheless, the randomization property of the XOR cellular automaton
(Theorem~\ref{thm:xor:randomization}) suggests a mathematically rigorous scenario
that makes Boltzmann's conclusion essentially true.
Indeed, the randomization implies that
for any finite set $A\subseteq\ZZ$,
the entropy $\hat{H}_A(T^t x)$ approaches (after ignoring a negligible set of time steps)
to its maximum value $\abs{A}\log 2$,
even if this convergence might be non-monotonic.\footnote{%
	Boltzmann's derivation for a system of particles was later made
	rigorous in a certain asymptotic regime (the Boltzmann-Grad limit)
	by the ground-breaking work of Lanford~\cite{Lan75} and others~\cite{CerIllPul94}.
}

For cellular automata, the uniform Bernoulli measure plays the role of
the Lebesgue measure on the phase space of a Hamiltonian system.
The analog of Liouville's theorem is the balance property,
which says that any surjective cellular automaton map $T:S^{\ZZ^d}\to S^{\ZZ^d}$
preserves the uniform Bernoulli measure.
Hence, Poincar\'e's theorem applies to all surjective cellular automata.
It says that starting from \emph{almost every} configuration $x$
(i.e., any configuration in a set of uniform Bernoulli measure~$1$),
every finite pattern occurring on $x$ (i.e., $q=x_A$ for some finite set $A\subseteq\ZZ^d$)
reappears on the same position 
infinitely many times (i.e., $(T^t x)_A=q$ for infinitely many time steps $t$).\footnote{%
	If the map $T$ is ergodic
	(e.g., if $T$ is the XOR map or XOR-transpose map),
	the average time between two consecutive reappearances is $2^{\abs{A}}$
	by Kac's recurrence theorem.
}
Note that this does not say anything about a starting configuration
that is \emph{not} typical for the uniform Bernoulli measure.

It is worth mentioning that surjective cellular automata
preserve the limit entropy per site,
that is, $h(T^t\pi)=h(\pi)$ for every macroscopic state $\pi$
and $t=1,2,\ldots$ (see e.g.~\cite{KarTaa14}).
On the other hand, randomization implies a jump at the limit to
$h(\mu)>\lim_{\substack{t\in J\\ t\to\infty}} h(T^t\pi)=h(\pi)$.
This may be understood as follows.\footnote{%
	For simplicity, we assume that $T$ has no non-trivial conserved quantity.
}
A sub-maximum value for $H_A(\pi)$ expresses the presence of correlations
among symbols with relative positions given by $A$.
The convergence of the entropy $H_A(T^t\pi)$ to its maximal value $\abs{A}\log\abs{S}$
for larger and larger finite sets $A\subseteq\ZZ^d$
indicates that the correlations are gradually distributed over larger and larger regions,
and are escaping to infinity as $t$ grows to infinity.

\section{How Far the Second Law Goes?}

The phenomenon described by the second law of thermodynamics
extends to all physical systems.
What constitutes a physical system and what is the exact statement of the second law
are much less clear.
Let us discuss few prerequisites for the presence of the randomization effect.
For simplicity, we focus on the case that
the cellular automaton has no non-trivial conserved quantity.

To fix the terminology, let us say that a cellular automaton $T:S^{\ZZ^d}\to S^{\ZZ^d}$
\emph{randomizes} a translation-invariant probability measure $\pi$ if
the Ces\`aro averages $\frac{1}{n}\sum_{t=0}^{n-1} T^t\pi$ converge weakly
to the uniform Bernoulli measure $\mu$ as $n$ goes to infinity.
Equivalently, $T$ randomizes $\pi$
if $T^t\pi\to\mu$ along a subsequence $J\subseteq\NN$ of density~$1$ of time steps
(see~\cite{JohRud95}, Corollary~1.4).

An obvious case in which randomization fails is when $T$ is not surjective.
Lack of surjectivity (or reversibility) has been suggested as
a mechanism behind the contrasting phenomenon of self-organization (see e.g.~\cite{Wol83}).
The requirement for $T$ to be surjective is a relaxation compared to reversibility
(let alone time-reversal symmetry), which is common among most microscopic physical theories.
Yet, surjectivity already guarantees the invariance of the uniform Bernoulli measure.
Moreover, surjective cellular automata are in some way close to being injective:
if two configurations $x,y$ differ on at most finitely many positions,
then $Tx$ and $Ty$ are distinct (see e.g.~\cite{Kar05}).\footnote{%
	A non-surjective cellular automaton may still act surjectively on a natural subspace
	(e.g., a mixing subshift of finite type).
	Randomization within such a subspace may still occur.
}

Besides surjectivity, the cellular automaton requires to have
certain degree of chaos in order for randomization to occur.
For instance, for a one-dimensional surjective cellular automata with equicontinuity points,\footnote{%
	A cellular automaton $T$ is \emph{equicontinuous} (or \emph{stable}) at a configuration $x$
	if for each finite set $A$, there is a finite set $B$ such that
	for any configuration $y$ that agrees with $x$ on $B$,
	$T^t x$ and $T^t y$ agree on $A$ for every $t\geq 0$.
	A cellular automaton with equicontinuity points is not sensitive, hence not chaotic (see~\cite{Kur03b}).
}
the Ces\`aro averages $\frac{1}{n}\sum_{t=0}^{n-1}T^t\pi$ with a Bernoulli starting measure $\pi$
converge but not necessarily to the uniform Bernoulli measure~\cite{BlaTis00}.
Such cellular automaton typically have too many distinct conserved quantities,
resulting in failure of any thermodynamic behaviour (see~\cite{Tak87,KarTaa14}).

Another obstacle for randomization is too much regularity in the starting configuration.
The simplest type of regularity is periodicity.
Note that a spatially periodic configuration is also temporally periodic.\footnote{%
	A configuration $x$ is said to be \emph{spatially periodic} if
	its translation orbit is finite, or equivalently, if there are
	$d$ linearly independent elements $q_1,q_2,\ldots,q_d\in\ZZ^d$
	such that $x_{a+n q_i} = x_a$ for all $a\in\ZZ^d$ and $n\in\ZZ$.
	A configuration $x$ is \emph{temporally periodic} if $T^p x=x$ for some $p>0$.
	Observe that every spatially periodic configuration is homogeneous.
}
Therefore, no cellular automaton can randomize a periodic configuration.
A spatially periodic configuration has zero entropy per site.
As a more sophisticated example of a regularity obstructing randomization,
consider the XOR cellular automaton.
The XOR cellular automaton has the following \emph{self-similarity} property,
which can be verified by induction:
$(T^{2^n}x)_i = x_i + x_{i + 2^n} \pmod{2}$ for every $n\geq 0$.
Define the \emph{duplicate} of a configuration $x$ as the configuration $Dx$
with $(Dx)_{2i}\isdef(Dx)_{2i+1}\isdef x_i$.
It follows from self-similarity
that $DTx=T^2 Dx$, or more generally $D^n Tx = T^{2^n}D^n x$.
For the uniform Bernoulli measure $\mu$, in particular, we find that
$T^{2^n}D^n\mu=D^n\mu$.
Note that if $\pi$ is a translation-ergodic measure,
the measure $\overline{D}\pi\isdef\frac{1}{2}(D\pi+\sigma D\pi)$
is also translation-ergodic and has entropy per site $\frac{1}{2}h(\pi)$.
Moreover, if $T^k\pi=\pi$, we get $T^{2k}\overline{D}\pi=\overline{D}\pi$.
Therefore, we have an infinite sequence $\overline{D}\mu,\overline{D}^2\mu,\overline{D}^3\mu,\ldots$
of distinct translation-ergodic measures (i.e., macroscopic states)
with positive entropy that are not randomized by the XOR cellular automaton.\footnote{%
	In fact, the XOR cellular automaton has a continuum of distinct translation-ergodic measures
	with positive entropy per site that are not randomized~\cite{Ein04}.
}

In summary, randomization is expected only if the cellular automaton
is surjective and ``sufficiently chaotic'', and
the starting configuration does not have ``too much regularity''.

\begin{acknowledgement}
	Research supported by ERC Advanced Grant 267356-VARIS
	of Frank den Hollander.
	I would like to thank Aernout van Enter, Nazim Fat\`es and Ville Salo
	for helpful comments and discussions.
	This article is dedicated with love and appreciation to
	my teacher Javaad Mesgari.
\end{acknowledgement}


\bibliographystyle{spmpsci}
\bibliography{bibliography}

\end{document}